\documentclass[12pt]{article}
\usepackage{cite}

\begin{document}

\begin{titlepage}

 \setcounter{page}{0}

 \begin{flushright}
  KEK-TH-1152 \\
  OIQP-07-07 \\
  YITP-07-29
 \end{flushright}

 \vskip 5mm
 \begin{center}
  {\Large\bf Fluxes of Higher-spin Currents and Hawking Radiations from
  Charged Black Holes \\}  
  \vskip 15mm

  {\large
  Satoshi Iso$^{1,}$\footnote{\tt satoshi.iso@kek.jp},
  Takeshi Morita$^{2,}$\footnote{\tt mtakeshi@yukawa.kyoto-u.ac.jp}
  and Hiroshi Umetsu$^{3,}$\footnote{\tt hiroshi\_umetsu@pref.okayama.jp}} 
  
  \vspace{5mm}
  $^1${\it Institute of Particle and Nuclear Studies \\
  High Energy Accelerator Research Organization (KEK) \\
  Oho 1-1, Tsukuba, Ibaraki 305-0801, Japan 
  }
 
  \vspace{5mm}

  $^2${\it Yukawa Institute for Theoretical Physics, Kyoto University
  \\
  Kyoto 606-8502, Japan}
  
  \vspace{5mm}
  $^3${\it Okayama Institute for Quantum Physics \\
  Kyoyama 1-9-1, Okayama 700-0015, Japan} 
 \end{center}

 \vskip 30mm 
 \centerline{{\bf{Abstract}}} 

 \vskip 3mm 
 This is an extended version of the previous paper (hep-th/0701272).  
 Quantum fields near horizons can be described in terms
 of an infinite set of two-dimensional conformal fields. We first
 generalize the method of Christensen and Fulling to charged black holes to
 derive fluxes of energy and charge.  These fluxes can be obtained by
 employing a conformal field theory technique. We then apply this technique
 to obtain the fluxes of higher-spin currents and show that the thermal
 distribution of Hawking radiation from a charged black hole can be
 completely reproduced by investigating transformation properties of the
 higher-spin currents under conformal and gauge transformations.

\end{titlepage}

\newpage

\setlength{\baselineskip}{6mm}

\section{Introduction}
\setcounter{footnote}{0}
\setcounter{equation}{0}

Hawking radiation is a characteristic quantum effect which arises in the
background space-time with event horizons \cite{Hawking1,Hawking2}.  The
radiation has the spectrum with the Planck distribution and it gives one of
the thermodynamic properties of black holes.  Hawking studied quantum
systems of matter in collapsing geometries and derived the thermal
radiation. Soon after Unruh revealed importance of choice of vacua,
i.e. choice of boundary conditions, and showed that eternal black holes also
emit the same radiation~\cite{Unruh}. Since then various derivations were
proposed, all of which lead to the same result.

Recently Robinson and Wilczek proposed a new derivation of Hawking
radiation~\cite{Robinson}. They realized that effective theories of matter
fields become two-dimensional and chiral near the event horizon of a
Schwarzschild black hole because of the structures proper to the
horizon. Then they showed that the existence of energy flux of the Hawking
radiation is required for the cancellation of the gravitational anomalies
near the horizon. The method was then generalized to charged black
holes~\cite{IUW1} by using the gauge anomaly in addition to the
gravitational anomaly and further to rotating black holes~\cite{IUW2,Murata}
and other various cases
\cite{Das,Setare,Xu,IMU1,Jiang,Jiang2,Jiang3,Kui,Shin,Peng,Jiang4,Chen,Murata2}.  In
these analyses, it is essential that each partial wave mode of quantum fields
near the horizon behaves as a two-dimensional field 
\cite{conformal}. Then, near the horizon,
outgoing modes are interpreted as right moving modes and ingoing modes as
left moving modes. Since the ingoing modes immediately fall into the
interior region of the black hole and never come out, they are considered to
be classically irrelevant to physics outside the black hole. However these
modes can not be neglected at the quantum level because if the ingoing modes
do not exist the theory becomes chiral and thus anomalous under the
diffeomorphism and gauge transformations. The fluxes of energy and charge
from the Hawking radiation are determined by vanishing conditions of the
anomalies at the horizon.

The method based on the gravitational and gauge anomalies successfully
reproduce the Hawking fluxes as mentioned above, but the fluxes of energy
and charge are merely a part of the information of the thermal
distribution. In Ref \cite{IMU2} we proposed a method using conformal field
theory technique to reconstruct the full thermal spectrum in a Schwarzschild
black hole. We there used the fact that each moment of the Planck
distribution, which is an integration of the distribution function
multiplied by an appropriate power of frequency, is equivalent to a flux of
the corresponding higher-spin current. The fluxes are given by expectation
values of the higher-spin currents which are defined with respect to the
coordinate system appropriate at infinity. These quantities at infinity are
related to the ones in the Kruskal coordinates by a conformal
transformation. We evaluated anomalous terms in transformation properties of
the higher-spin currents under the conformal transformation. These terms
give main contributions to the fluxes. The fluxes are determined by imposing
the regularity condition that the expectation values of the currents should
be regular at the horizon in the Kruskal coordinates.

In this paper we generalize the method to charged black holes by considering
a gauge transformation in addition to the conformal transformation.  In a
gauge that the background $U(1)$ gauge field vanishes at infinity, the gauge
potential is singular at the horizon. In order to adequately impose the
regularity condition on quantities at the horizon, we have to use a gauge in
which the background field behaves regularly there. Hence we need to know
the transformation properties of the higher-spin currents under suitable
gauge and conformal transformations, which map physical quantities near the
horizon to those at infinity. We will derive the fluxes of the higher-spin
currents containing the complete information of the thermal spectrum of
the Hawking radiation by using such relations and the regularity condition.

This paper is organized as follows. In section \ref{GCF} we make a
generalization of the method by Christensen and Fulling~\cite{CF} to charged
black holes. Fluxes of energy and charge are obtained by solving
conservation equations of energy-momentum tensor and $U(1)$ current and the
anomaly equations for conformal and chiral symmetries. 
In section \ref{current-EM}, we derive the same fluxes by using conformal
field theory technique. This analysis is applied to higher-spin currents to
derive their fluxes. We conclude in section \ref{conclusion}.

\section{Christensen-Fulling method for charged black holes \label{GCF}}
\setcounter{equation}{0}

Christensen and Fulling obtained the flux of Hawking radiation 
from neutral black holes by solving the conservation equations
of the energy-momentum tensor together with the information of 
the trace anomaly~\cite{CF}. In their seminal paper, they showed that 
the outgoing flux at infinity cannot be put zero if we require
(1) conservation of energy-momentum tensor (2) trace anomaly
and (3) regularity at the future horizon.
Their calculation works well in the two-dimensional case but
in higher dimensions, since the number of components of 
the energy-momentum tensor is larger than the above three
information, all the components cannot be determined fully 
without further information.
Also a generalization to charged (or rotating) 
black holes  needs further information to determine the flux. 
In this section, we generalize the method by Christensen 
and Fulling to two-dimensional charged black holes
by  further considering the conservation of the gauge current
and the chiral anomaly equation. 
\setcounter{equation}{0}
\subsection{Reissner-Nordstr\"om black hole}
We first summarize the basics of Reissner-Nordstr\"om black holes.
The metric and the gauge potential of Reissner-Nordstr\"om black holes
with mass $M$ and charge $Q$ are given by 
\begin{eqnarray}
 ds^2 &=& f(r)dt^2-\frac{1}{f(r)}dr^2-r^2d\Omega_2^2,\\
 A_t &=& -\frac{Q}{r}, \label{bg-gauge}
\end{eqnarray}
where 
\begin{equation}
 f(r) = 1-\frac{2M}{r}+\frac{Q^2}{r^2} = \frac{(r-r_+)(r-r_-)}{r^2}
 \end{equation}
 and the radius of outer (inner) horizon $r_\pm$ is given by
 \begin{equation}
 r_\pm = M \pm \sqrt{M^2-Q^2}.
\end{equation}
It is useful to define the  tortoise coordinate 
by solving $  dr_* = dr/f $  as
\begin{equation}
  r_* = r + \frac{1}{2\kappa_+}\ln \frac{|r-r_+|}{r_+}
  + \frac{1}{2\kappa_-}\ln \frac{|r-r_-|}{r_-}.
  \end{equation}
  Here the surface gravity at $r_\pm$ is given by
  \begin{equation}
  \kappa_\pm = \frac{1}{2}f'(r_\pm) = \frac{r_\pm - r_\mp}{2r_\pm^2}.
\end{equation}
In the following we consider near the outer horizon.
First we define the light-cone coordinates, 
$u=t-r_*$ and $v=t+r_*$. $u(v)$ are the 
outgoing (ingoing) coordinates and the metric 
in these coordinates becomes as
\begin{equation}
 ds^2 = f(dt^2 - dr_*^2) - r^2d\Omega^2 
 = fdudv - r^2d\Omega^2. 
 \label{uv}
\end{equation}
In order to investigate the physics near the outer horizon,
since $(u,v)$ coordinate is still singular at the horizon, 
it is important to introduce a regular coordinate,  
Kruskal coordinate, defined by the transformations
\begin{equation}
   U=-e^{-\kappa_+ u}, \qquad V=e^{\kappa_+ v}.
   \label{UVtransf}
\end{equation}
The metric becomes
\begin{equation} 
 ds^2  = \frac{r_+r_-}{\kappa_+^2}\frac{1}{r^2}e^{-2\kappa_+ r}
       \left(\frac{r_-}{r-r_-}\right)^{\frac{\kappa_+}{\kappa_-}-1}
       dUdV  - r^2d\Omega^2.
       \label{UV}
\end{equation}
If we restrict to see the two-dimensional $(r,t)$ section,
both of these coordinates (\ref{uv}), (\ref{UV}) have the forms of 
the conformal gauge
\begin{eqnarray}
 ds^2 = e^{\varphi(u,v)} dudv = e^{\varphi'(U,V)}dUdV.
\end{eqnarray}
The transformation (\ref{UVtransf}) is a conformal 
transformation from an asymptotically flat coordinate
to a regular coordinate near the horizon.

Hereafter we omit the subscript $+$ of the surface gravity $\kappa_+$ at the
outer horizon because $\kappa_-$ does not appear in the following analysis.

\subsection{Solving conservation equations}
We now solve conservation equations for the gauge current and the
energy-momentum tensor together with the information of trace and chiral
anomalies.  Here we consider a charged matter field in two-dimensional black
holes.  If we get a two-dimensional system by a dimensional reduction of
higher-dimensional ones, we need to take care of the effects of angular
components of the gauge current or the energy-momentum tensor.  In the
following, we simply neglect these effects and consider purely
two-dimensional cases for simplicity.  Then conservation laws for the gauge
current and the energy-momentum tensor in the background of gravitational
and gauge fields are given by
\begin{eqnarray}
 \label{cons-EM}
  \nabla_\mu T^\mu_\nu &=& F_{\mu\nu} J^\mu, \\
  \label{cons-vec}
  \nabla_\mu J^\mu &=& 0.
\end{eqnarray}
If we consider conformal matter fields in this background, 
the energy-momentum tensor has the trace anomaly
\begin{equation}
 \label{trace}
  T^\mu_\mu = \frac{c}{24\pi} R, 
  \end{equation}
where $c$ is the central charge of the conformal fields, 
$R=-4e^{-\varphi}\partial_u\partial_v \varphi$ is a scalar curvature 
and the conformal factor $\varphi$ in the $(u,v)$ coordinate is given by
$\varphi = \ln f(r)$. The two-dimensional chiral current, which
is related to the gauge current by $J^{5\mu} = \epsilon^{\mu\nu} J_\nu$, 
has an anomaly. Here $\epsilon^{\mu\nu}$ is a covariant antisymmetric
tensor, $\epsilon^{uv}=2e^{-\varphi}$. For a fermion field with charge $e$,
it is given by 
\begin{equation}
 \label{cons-axial}
  \nabla_\mu J^{5\mu} = \frac{e^2}{2\pi}\epsilon^{\mu\nu}F_{\mu\nu}.
\end{equation}
For more general cases, the coefficients depend on the matter contents
but in the following we absorb them  into the charge $e$ and use
the same conservation equation (\ref{cons-axial}).

First let us solve the conservation equations for gauge (\ref{cons-vec})
and chiral currents
(\ref{cons-axial}).
By using the relation $J^{5\mu} = \epsilon^{\mu\nu} J_\nu$, 
these two equations can be solved as
\begin{eqnarray}
 && J_u= j(u) + \frac{e^2}{\pi}A_u, \qquad 
 J_v=  \tilde{j}(v) + \frac{e^2}{\pi}A_v.
  \label{solvecurrent}
\end{eqnarray}
where $j(u)$ (or $\tilde{j}(v)$) is a holomorphic (or anti-holomorphic)
function satisfying $\partial_v j(u) = \partial_u \tilde{j}(v) = 0$.
As we see in the next section, 
these (anti-) holomorphic currents play important roles
in the conformal field theories.
Note that they are different from the ordinary currents $J_u, J_v$
by a term proportional to the background gauge potential
and not invariant under gauge transformations.

In order to determine these functions in the black hole background,
we use the same boundary conditions as those used in the
paper by Christensen and Fulling.
Namely, for an outgoing flux $j(u)$, we impose that the 
physics should be regular in the Kruskal coordinate $U$.
Since $J_U=J_u/(-\kappa U)$, the flux $J_u$ must vanish at the future outer 
horizon $U=0$.  Thus
 $j(u)$ is determined to be
\begin{eqnarray}
 \label{vev-current}
 j(u) = -\frac{e^2}{\pi} A_u(r_+).
\end{eqnarray} 
The anti-holomorphic part 
$\tilde{j}(v)$ is determined by the condition that there is no ingoing flux
from infinity ($r\rightarrow\infty$) and given by
\begin{eqnarray}
 \tilde{j}(v) = 0.
\end{eqnarray}
These two conditions correspond to taking the so-called Unruh vacuum.
Hence the radial component of the electromagnetic current in the charged
black hole is given by
\begin{eqnarray}
 J^r = J_u - J_v
  = -\frac{e^2}{\pi} A_u(r_+)
  = \frac{e^2 Q}{2\pi r_+}.
\end{eqnarray}

Similarly the expectation value of the energy-momentum tensor
in the charged black hole can be determined by solving
eqs.(\ref{cons-EM}) and (\ref{trace}).
The $uu$ and $vv$ components of them can be 
solved as 
\begin{eqnarray}
 T_{uu} &=& t(u) + \frac{c}{24\pi}
  \left(\partial_u^2 \varphi - \frac{1}{2}(\partial_u \varphi)^2\right)
  + \frac{e^2}{\pi} A_u^2 + 2 A_u j(u),  \label{solveemtensor}  \\ 
 T_{vv} &=&  \tilde{t}(v)  + \frac{c}{24\pi}
  \left(\partial_v^2 \varphi - \frac{1}{2}(\partial_v \varphi)^2\right)
  + \frac{e^2}{\pi} A_v^2 + 2 A_v \tilde{j}(v).
\end{eqnarray}
Here $t(u)$ is a holomorphic function and $\tilde{t}(v)$ is an
anti-holomorphic function, which can be determined by the following boundary
conditions.

Imposing the regularity condition at the outer horizon, 
$T_{uu}$ must vanish at the future outer horizon $U=0$.
Hence $t(u)$ is determined as
\begin{eqnarray}
 \label{vev-em}
 t(u) = \frac{c}{192\pi}\left(f'(r_+)\right)^2 
  + \frac{e^2}{\pi} A_u^2(r_+).
\end{eqnarray}
Similarly 
$\tilde{t}(v)$ is determined by requiring that 
there is no ingoing flux from infinity as
\begin{eqnarray}
 \tilde{t}(v) = 0.
\end{eqnarray}
Hence the $rt$-component of the energy-momentum tensor $T^r_t$ is obtained as
\begin{eqnarray}
 T^r_t = T_{uu} - T_{vv}
  = -\frac{e^2}{2\pi}A_t(r_+)A_t(r) 
  + \frac{c}{192\pi}\left(f'(r_+)\right)^2 
  + \frac{e^2}{4\pi} A_t^2(r_+).
\end{eqnarray}
The asymptotic flux of the energy is
given by the asymptotic value of $T^r_t$ at 
$r\rightarrow\infty$,
\begin{eqnarray}
 T^r_t \longrightarrow \frac{c}{192\pi}\left(f'(r_+)\right)^2 
  + \frac{e^2}{4\pi} A_t^2(r_+).
\end{eqnarray}
The flux reproduces the correct flux of Hawking radiation 
from the charged black hole~\cite{IUW1}.

\section{$\mathbf{U(1)}$ current and energy-momentum tensor \label{current-EM}}
\setcounter{equation}{0}

In section \ref{GCF}, we generalized the method by Christensen and Fulling to
charged black holes and obtained the fluxes of charge and energy.  In this
section we derive the same result by using a technique of conformal field
theory without resorting to the covariant calculation done in section
\ref{GCF}. Here we consider a charged fermionic field in the black hole
background. It is also easy to apply the method to a bosonic field.

We define the holomorphic $U(1)$ current 
$:\psi^\dagger(u) \psi(u):$ 
by using the point splitting regularization as follows ~\cite{yellow},
\begin{eqnarray}
 :\psi^\dagger(u) \psi(u): \equiv 
  \lim_{\epsilon \rightarrow 0} 
  \left(\psi^\dagger(u+\epsilon) \psi(u) 
   + \frac{i}{2\pi \epsilon}\right).
\end{eqnarray}
Here we used $\psi(z) \psi^\dagger(w) \sim -i/(2\pi (z-w))$.  Under a
holomorphic $U(1)$ transformation\footnote{ 
A holomorphic transformation $\delta_h$ with an infinitesimal parameter
$\lambda(u)$ is a combination of a gauge transformation $\delta_G$ and a
chiral transformation $\delta_C$ with the same parameter $\lambda(u)$;
$\delta_h = \frac{1}{2}(\delta_G - \delta_C)$.}, 
the fermion transforms as
$\psi^{(\Lambda)}(u) = e^{ie\Lambda(u)}\psi(u)$.  
Then the transformation law for the current can be calculated as
\begin{eqnarray}
 :{\psi^{(\Lambda)}}^\dagger(u)\psi^{(\Lambda)}(u):
  &=& \lim_{\epsilon\rightarrow 0}
  \left(
   e^{-ie\Lambda(u+\epsilon) + ie\Lambda(u)}\psi^\dagger(u+\epsilon)\psi(u)
   + \frac{i}{2\pi\epsilon}
  \right) \nonumber \\
  &=& \lim_{\epsilon\rightarrow 0}
  \left[
   e^{-ie\Lambda(u+\epsilon) + ie\Lambda(u)}
   \left(:\psi^\dagger(u+\epsilon)\psi(u): - \frac{i}{2\pi\epsilon}\right)
   + \frac{i}{2\pi\epsilon}
  \right] \nonumber \\
  &=& :\psi^\dagger(u)\psi(u): - \frac{e}{2\pi}\partial_u \Lambda(u).
  \label{currentgaugetr}
\end{eqnarray}
Note that, since we have defined the current by the point splitting method
without introducing a Wilson line, the current is not invariant under the
gauge transformation. The current is identified with the holomorphic current
$j(u)$ defined in (\ref{solvecurrent}), which is also not gauge invariant,
as $j(u) = e :\psi^\dagger(u)\psi(u):$. 

Under a conformal transformation $u\rightarrow w(u)$,
the fermion $\psi(u)$ transforms as  
$\psi(u)=(\partial_u w(u))^{1/2}\psi^{(w)}(w(u))$.
Hence  the current transforms as a tensor with a conformal weight 1 as
expected;
\begin{eqnarray}
 :\psi^\dagger(u) \psi(u): 
  = \partial_u w(u) :{\psi^{(w)}}^\dagger (w(u)) \psi^{(w)}(w(u)):.
  \label{currentconftr}
\end{eqnarray}

We apply these transformation properties
 (\ref{currentgaugetr}) and (\ref{currentconftr})
to the charged black hole background.
Unruh vacuum is defined by using modes associated with the Kruskal
coordinate $U=-e^{-\kappa u}$, which is regular at the outer horizon.
(i.e. Unruh vacuum is annihilated by positive  frequency modes
defined in terms of the Kruskal coordinate.)   
Furthermore 
the background gauge potential (\ref{bg-gauge}) is singular at the outer
horizon in the Kruskal coordinates because 
$A_U = (-1/\kappa U) A_u = (-1/2\kappa U) A_t$, $(A_r = 0)$. 
In order to impose boundary conditions for currents, 
we need to take  a gauge which is regular at the outer horizon 
$U=0$ in the Kruskal coordinates.
Such a gauge choice is given by $A_t = -Q/r + Q/r_+$.
We denote the gauge fields in two different gauges as
\begin{eqnarray}
 A_t^{(u)} = -\frac{Q}{r}, \qquad
  A_t^{(U)} = -\frac{Q}{r} + \frac{Q}{r_+}.
\end{eqnarray}
The first gauge potential vanishes at $r\rightarrow \infty$ and it is
appropriate to use it to measure the frequency at an asymptotic
infinity. Hence we use the superscript $(u)$.  The second one does not
satisfy this property but it vanishes at the outer horizon and the
corresponding $A_U$ behaves regularly there. This is why we use the
superscript $(U)$ for this gauge.  They are related by a gauge
transformation with a parameter $tQ/r_+$; $A_t^{(U)} = A^{(u)}_t +
\partial_t (tQ/r_+)$.  In the holomorphic sector, we consider a holomorphic
$U(1)$ transformation with a parameter $\Lambda(u) = uQ/r_+$
to implement this gauge transformation\footnote{ 
Our choice of $\Lambda(u)$ seems to be twice as large as the one
required by the gauge transformation since
$\frac{tQ}{r_+} = \frac{uQ}{2r_+} + \frac{vQ}{2r_+}$. 
This factor of two arises from a relation between the fermionic field used
to construct the conformal current and the original fermionic field
in the gauge field background. We can show that by taking this field
redefinition into account the relations between the covariant and conformal
currents (eqs.(\ref{solvecurrent}) or (\ref{solveemtensor})) can be
correctly obtained. We will give further discussions on this issue in our
future publication \cite{IMU3}.
}.
Hereafter we explicitly indicate the gauge dependence of expectation values
of (gauge-dependent) operators by using the subscripts $A^{(u)}$ or
${A^{(U)}}$ as $\langle {\cal O} \rangle_{A^{(u)}}$ and $\langle {\cal O}
\rangle_{A^{(U)}}$.

First the expectation values of the current in these gauges,
$\langle :\psi^\dagger(u) \psi(u): \rangle_{A^{(u)}}$ 
and
$\langle :{\psi^{(U)}}^\dagger(U) \psi^{(U)}(U): \rangle_{A^{(U)}}$,
are related as follows; 
\begin{eqnarray}
 \langle :\psi^\dagger(u) \psi(u): \rangle_{A^{(u)}}
  &=& 
  \langle :{\psi}^\dagger(u) \psi(u): \rangle_{A^{(U)}}
  + \frac{e}{2\pi}\partial_u \Lambda(u)
  \nonumber \\
  &=& 
  -\kappa U 
  \langle :{\psi^{(U)}}^\dagger(U) \psi^{(U)}(U): \rangle_{A^{(U)}}
  + \frac{e}{2\pi}\partial_u \Lambda(u).
\end{eqnarray} 
In the second line we make a conformal transformation
from $u$ coordinate to $U$. 
We impose the boundary condition that physical quantities 
should behave regularly at the outer future horizon $U=0$ in the Kruskal 
coordinate $U$ and the gauge $A^{(U)}$ which is regular there.
In other words,  
$\langle :{\psi^{(U)}}^\dagger(U) \psi^{(U)}(U): \rangle_{A^{(U)}}$ 
should be finite at $U=0$.
This condition determines  the flux of the charge current at infinity 
to be
\begin{eqnarray}
 \langle :\psi^\dagger(u) \psi(u): \rangle_{A^{(u)}}
  = \frac{eQ}{2\pi r_+},
\end{eqnarray} 
where we have used $\Lambda(u) = uQ/r_+$. This coincides with
(\ref{vev-current}) obtained by solving the conservation equations.

The energy-momentum tensor is defined similarly by the 
following point splitting method,
\begin{eqnarray}
 && :\frac{i}{2}\left(\psi^\dagger(u)\partial_u \psi(u) 
  - \partial_u \psi^\dagger(u) \psi(u)\right): \nonumber \\ 
 && \qquad
 \equiv \lim_{\epsilon\rightarrow 0}
  \left[
   \frac{i}{2}\left(\psi^\dagger(u+\epsilon)\partial_u \psi(u) 
   - \partial_u \psi^\dagger(u+\epsilon) \psi(u) \right)
   - \frac{1}{2\pi\epsilon^2}
  \right].
  \label{confemtensor}
\end{eqnarray}
Under a conformal transformation, the energy-momentum tensor transforms as
\begin{eqnarray}
 && :\frac{i}{2}\left(\psi^\dagger(u)\partial_u \psi(u) 
  - \partial_u \psi^\dagger(u) \psi(u)\right): \nonumber \\
 && \quad 
  = \left(\partial_u w(u)\right)^2 
  :\frac{i}{2}\left({\psi^{(w)}}^\dagger(w(u))\partial_w \psi^{(w)}(w(u)) 
  - \partial_w {\psi^{(w)}}^\dagger(w(u)) \psi^{(w)}(w(u))\right): 
  \nonumber \\
  && \qquad - \frac{1}{24\pi} \left\{w, u\right\}^{(1)}, 
 \label{conftremtensor}
\end{eqnarray}
where $\{w, u\}^{(1)}$ is the Schwarzian derivative
\begin{eqnarray}
 \{w, u\}^{(1)} = \frac{\partial_u^3 w(u)}{\partial_u w(u)}
  -\frac{3}{2}\left(\frac{\partial_u^2 w(u)}{\partial_u w(u)}\right)^2.
\end{eqnarray}
This energy-momentum tensor corresponds to the holomorphic tensor  $t(u)$
defined in (\ref{solveemtensor}). The extra factor in the transformation 
(Schwarzian derivative) comes from the Weyl transformation of
the second term in the right hand side of (\ref{solveemtensor}).

Under the $U(1)$ transformation, the energy-momentum tensor
(\ref{confemtensor}) transforms as
\begin{eqnarray}
 && :\frac{i}{2}
  \left(
   {\psi^{(\Lambda)}}^\dagger(w(u))\partial_u \psi^{(\Lambda)}(u) 
  - \partial_u {\psi^{(\Lambda)}}^\dagger(u) \psi^{(\Lambda)}(u)
  \right):
  \nonumber \\
  && \qquad 
   = :\frac{i}{2}\left(\psi^\dagger(u)\partial_u \psi(u) 
   - \partial_u \psi^\dagger(u) \psi(u)\right):
   -e\partial_u \Lambda (u) :\psi^\dagger(u)\psi(u):
   \nonumber \\
 && \qquad \quad
   + \frac{e^2}{4\pi}\left(\partial_u \Lambda(u)\right)^2.
   \label{gaugetremtensor}
\end{eqnarray}
This is not gauge invariant, which can be understood
either from the gauge dependent factors in (\ref{solveemtensor})
or from the point splitting definition without a Wilson line 
(\ref{confemtensor}).

Now we apply these transformations (\ref{conftremtensor}) and
(\ref{gaugetremtensor}) to derive the expectation values of 
the energy-momentum tensor in 
the charged black hole background.
The energy-momentum tensor in the $u$ coordinate with 
the gauge $A^{(u)}$ is related to that in the $U$ coordinate
with the gauge $A^{(U)}$ as
\begin{eqnarray}
 && \langle :\frac{i}{2}\left(\psi^\dagger(u)\partial_u \psi(u) 
  - \partial_u \psi^\dagger(u) \psi(u)\right): \rangle_{A^{(u)}} 
  \nonumber \\
  && \quad
   = (\kappa U)^2 \langle 
   :\frac{i}{2}\left({\psi^{(U)}}^\dagger(U)\partial_U \psi^{(U)}(U) 
   - \partial_U {\psi^{(U)}}^\dagger(U) \psi^{(U)}(U)\right): 
   \rangle_{A^{(U)}}
   \nonumber \\
 && \qquad
   + e \partial_u \Lambda(u) (\kappa U)
   \langle :{\psi^{(U)}}^\dagger(U) \psi^{(U)}(U): \rangle_{A^{(U)}}
  -\frac{1}{24\pi} \{U, u\}
  + \frac{e^2}{4\pi}\left(\partial_u \Lambda(u)\right)^2.
\end{eqnarray}
Hence, by imposing the regularity condition for
$T_{UU}$ in the $A^{(U)}$ gauge, we have
\begin{eqnarray}
 \langle :\frac{i}{2}\left(\psi^\dagger(u)\partial_u \psi(u) 
  - \partial_u \psi^\dagger(u) \psi(u)\right): \rangle_{A^{(u)}} 
  = \frac{\kappa^2}{48\pi} + \frac{e^2Q^2}{4\pi r_+^2}.
\end{eqnarray}
This value is equal to (\ref{vev-em}) in the previous section.

\section{Higher-spin currents \label{HS}}
\setcounter{equation}{0} 

We now generalize the method in the previous section to higher-spin
currents. The fluxes of these currents correspond to higher moments of the
frequency integral of the Hawking radiation.  Hence, if the fluxes of all 
types of higher-spin currents are obtained, we can fully reproduce the
thermal spectrum of Hawking radiations from (charged) black holes.

In order to get the fluxes of charge and energy flows, we used a covariant
approach in section \ref{GCF} and a conformal field theory approach in
section \ref{current-EM}.  As we saw, they are equivalent but the latter
approach is much simpler since we do not need either to know the covariant
formulation of the (trace or chiral) anomaly equations or to solve the
conservation equations of currents. But there is an issue which should be
clarified to generalize the above approach to cases of higher-spin
currents. We need to know relations between covariant higher-spin currents
and corresponding (anti-)holomorphic quantities like (\ref{solvecurrent})
and ({\ref{solveemtensor}}) in order to show that expectation values derived
by using conformal field theory technique adopted here coincide with fluxes
obtained by covariant calculation. In this paper, we use the fact,
sufficient for the present analysis, that differences between covariant
higher-spin currents and holomorphic ones approach to 0 at infinity. Then we
show that fluxes derived by conformal field theory approach coincide with
certain moments of the Fermi-Dirac distribution of the Hawking radiations,
which will be defined below. We will give full discussions on this issue in 
our future publication~\cite{IMU3}.

The holomorphic component of the $(n+1)$-th rank higher-spin current is
given by a linear combination of 
$\partial_u^m \psi^\dagger \partial_u^{n-m}\psi$.
Since we can show that all these terms give the same contribution to 
an expectation value of the current in the black hole
background~\cite{IMU2}, we here consider only 
$\psi^\dagger \partial_u^n\psi$. 
First we introduce higher-spin currents which are regularized 
by the point splitting as follows,
\begin{eqnarray}
:\psi^\dagger (u) \partial_u^n \psi(u): 
 \equiv \lim_{\epsilon\rightarrow 0} 
 \left[\psi^\dagger (u+\epsilon) \partial_u^n \psi(u) 
  + \frac{i n!}{2\pi\epsilon^{n+1}}\right].
 \end{eqnarray}
It is convenient to introduce
a generating function of the higher-spin currents,
\begin{eqnarray}
 :\psi^\dagger (u) \psi(u+a): 
  \equiv \sum_{n=0}^\infty \frac{a^n}{n!} 
  :\psi^\dagger (u)\partial_u^n \psi(u):. 
  \label{higherspincurrents}
\end{eqnarray}
This generating function is defined by the right hand side and should be
understood as a formal power series of $a$ whose coefficients are normal
ordered operators located at $u$.
The fermion has a conformal weight 1/2 and 
$\psi$ transforms as 
$\psi(u)=(\partial_u w(u))^{1/2}\psi^{(w)}(w(u))$ under a conformal
transformation $u\rightarrow w(u)$. 
Therefore the generating function transforms as 
\begin{eqnarray}
 \label{HS-conf}
 :\psi^\dagger (u) \psi(u+a): &=& 
  \psi^\dagger(u)\psi(u+a) - \frac{i}{2\pi a} \nonumber \\
  &=& \left[\partial_u w(u) \partial_u w(u+a)\right]^{1/2}
   :\psi^{(w)\dagger} (w(u)) \psi^{(w)}(w(u+a)): \nonumber \\
   && + A_f (w,u).
\end{eqnarray}
Here $A_f(w, u)$ is a generating function of generalized Schwarzian
derivatives for higher-spin currents (\ref{higherspincurrents})
and given by
\begin{eqnarray}
 A_f(w, u) = \sum_{n=0}^\infty \frac{(-ia)^n}{n!} \{w,u \}^{(n)} =
 \frac{i}{2\pi}
 \left(
  \frac{\left[\partial_u w(u)\partial_u w(u+a)\right]^{1/2}}{w(u+a)-w(u)}
  -\frac{1}{a}
  \right).
\end{eqnarray}
$\{w,u\}^{(1)}$ is proportional to the ordinary Schwarzian derivative.

Next under the  $U(1)$ transformation
$\psi^{(\Lambda)}(u) = e^{ie\Lambda(u)}\psi(u)$,
the generating function transforms as
\begin{eqnarray}
 \label{HS-gauge}
 :{\psi^{(\Lambda)}}^\dagger(u)\psi^{(\Lambda)}(u+a):
  &=& e^{ie(\Lambda(u+a) - \Lambda(u))} :\psi^\dagger(u)\psi(u+a):
  \nonumber \\
  && - \frac{i}{2\pi a} \left(1-e^{ie(\Lambda(u+a) - \Lambda(u))}\right).
\end{eqnarray}
As we saw in the simplest case, the inhomogeneous term arises
because of the point splitting regularization without a Wilson line.

Similarly to the cases of the $U(1)$ current and energy-momentum tensor, we
apply these transformation properties (\ref{HS-conf}) and (\ref{HS-gauge}) 
to derive the expectation values of the higher-spin currents in the black
hole background. From these equations, the expectation values of the
generating function in the $u$ coordinate with the gauge $A^{(u)}$ is related
with that in the $U$ coordinate with the gauge $A^{(U)}$ as follows,
\begin{eqnarray}
 && \langle :\psi^\dagger(u) \psi(u+a): \rangle_{A^{(u)}}
  \nonumber \\
 && \qquad
  = e^{-ie(\Lambda(u+a) - \Lambda(u))} 
  \left[
   (\kappa U) e^{-\kappa a/2} 
   \langle :{\psi^{(U)}}^\dagger(U(u)) \psi^{(U)}(U(u+a)): \rangle_{A^{(U)}}
   + A_f (U,u)
  \right]
  \nonumber \\
 &&  \qquad \quad
  - \frac{i}{2\pi a} \left(1 - e^{-ie(\Lambda(u+a) - \Lambda(u))} \right),
\end{eqnarray}
where the explicit form of $A_f(U, u)$ is
\begin{equation}
 A_f(U, u) = \frac{i}{2\pi a}
  \left(\frac{\kappa a}{2}\frac{1}{\sinh\frac{\kappa a}{2}} - 1\right).
\end{equation}
By imposing the regularity condition at the horizon $U=0$, the fluxes at 
infinity are determined as
\begin{eqnarray}
 \label{vev-genfun}
 \langle :\psi^\dagger(u) \psi(u+a): \rangle_{A^{(u)}} 
  = \frac{i}{2\pi a} 
  \left(
   e^{-i\frac{eQa}{r_+}}
  \frac{\kappa a}{2}\frac{1}{\sinh \frac{\kappa a}{2}}  - 1
  \right).
\end{eqnarray}
This can be interpreted as the temperature-dependent part of a finite Green
function for a charged fermion~\cite{Gibbons-Perry}, as mentioned in Ref
\cite{IMU2} in the case of Schwarzschild black holes. By expanding this as a
power series of $a$, the fluxes of the higher-spin currents are obtained as
\begin{eqnarray}
 \label{vev-hs}
 && \langle : i^n \psi^\dagger(u)\partial_u^n \psi(u): \rangle_{A^{(u)}}
 \nonumber \\
 && \hspace{10mm}
  = \sum_{m=1}^{\lceil \frac{n}{2} \rceil} 
  \left(\frac{eQ}{r_+}\right)^{n+1-2m}
  \frac{n!\left(1-2^{1-2m}\right) B_m \kappa^{2m}}{2\pi (2m)!(n-2m+1)!}
  + \frac{1}{2\pi (n+1)} \left(\frac{eQ}{r_+}\right)^{n+1}. 
\end{eqnarray}  
Here $B_m$ is the Bernoulli number $(B_1 = 1/6, ~B_2 = 1/30)$ and 
$\lceil x \rceil$ is the ceiling function, which returns the smallest
integer not less than $x$. 

It can be seen that these expectation values correspond to moments of the
Fermi-Dirac distribution\footnote{
Contributions from anti-holomorphic parts of the higher-spin currents to the
fluxes vanish due to the boundary condition corresponding to the Unruh
vacuum, i.e. no ingoing flux at infinity.
}. 
Thermal radiations of a fermion with charge $q$
from Reissner-Nordstr\"om black holes satisfy the Fermi-Dirac distribution
$N_q(\omega)$ with a chemical potential corresponding to the value of the
electric potential at the horizon,
\begin{equation}
 N_q(\omega) = \frac{1}{e^{\beta (\omega - \frac{qQ}{r_+})} + 1}. 
\end{equation}
The flux of the higher-spin current (\ref{vev-hs}) is equal to the following
quantity
\begin{eqnarray}
 \int_0^\infty \frac{d\omega}{2\pi}
  \left[\omega^n N_e(w) - (-\omega)^n N_{-e}(w)\right].
\end{eqnarray}
This can be interpreted as a moment of the thermal flux which consists of 
contributions from the fermion with charge $e$ and its antiparticle with
charge $-e$. Therefore one can reconstruct the whole information of the
thermal radiation from the fluxes of the higher-spin currents derived in
this section by using conformal field theory technique.


\section{Conclusion and discussion \label{conclusion}}
\setcounter{equation}{0}

In this paper, we first generalized the method by Christensen and Fulling to
charged black holes, and derived the fluxes of energy and charge by solving
the conservation equations and anomaly equations.  We then employed the
conformal field theory method, used for Schwarzschild black holes in
Ref.\cite{IMU2}, to charged black holes, and obtained these fluxes by
investigating transformation properties of the gauge and energy-momentum
currents under conformal and gauge transformations.  It was crucial to
consider the transformations from $(u, A^{(u)})$ to $(U, A^{(U)})$, where
the former are suitable coordinate and gauge potential to evaluate
quantities at infinity while the latter are suitable at the horizon.
Finally we applied this method to higher-spin currents and calculated their
expectation values in the charged black hole background.  We showed that the
expectation value of each higher-spin current in the Unruh vacuum exactly
coincides with the corresponding specific moment of the thermal
distribution.

The anomaly method has been  universally applied to any
type of black holes, black rings or cosmological models with horizons
\cite{Robinson,IUW1,IUW2,Murata,Das,Setare,Xu,IMU1,Jiang,Jiang2,Jiang3,Kui,Shin,Peng,Jiang4,Chen,Murata2}. 
The essence is that the effective theories near horizons can be described by
two-dimensional conformal fields in gravitational or gauge field
backgrounds. The gauge field is either the original $U(1)$ gauge field or an 
effective gauge field induced by a dimensional reduction of
higher-dimensional metric fields.  Since our present analysis is also based
on the same property near the horizon, the analysis of the higher-spin
currents and derivations of the thermal radiations 
can be similarly applied
to all kinds of black holes, black rings or cosmological models.

While we here used a conformal field theory technique, it would be possible
to generalize the covariant calculation given in section \ref{GCF} to the
cases of higher-spin currents. To do this, it is necessary to construct
covariant totally symmetric traceless higher-spin currents and to calculate
the trace anomalies of these currents. Then the fluxes of these currents can
be obtained by solving conservation equations and anomaly equations of the
currents. In this analysis, relations between the covariant and conformal
higher-spin currents, which are generalizations of eqs. (\ref{solvecurrent})
and (\ref{solveemtensor}), will be also obtained.  We will discuss them in
our future publication~\cite{IMU3}.

The derivation of Hawking radiations based on the anomaly method implies
that the radiation is determined by the information of backgrounds at the
horizon, not by the asymptotic charges measured at infinity.  It
would be interesting to see how a background which damps rapidly at
infinity can affect the radiation. 

\section*{Acknowledgements}
We would like to thank H. Kodama and Y. Nakayama
for useful information.  We also thank
A. Dhar, G. Mandal, S. R. Wadia and  other members of the Tata Institute
of Fundamental Research for their hospitality and stimulation during our
stay. 


\end{document}